\documentclass[]{eps}

\usepackage{graphicx,natbib,amssymb}
\usepackage{times}
\newcommand{\mdot}{\mbox{$M_{\odot}$ yr$^{-1}$}}
\newcommand{\msun}{$M_{\odot}$}
\def\aap{Astron.~Astrophys.}                
\def\aj{Astron.~J.}                   
\def\apj{Astrophys.~J.}                 
\def\apjl{Astrophys.~J.}                
\def\apjs{Astrophys.~J.~Suppl.}               
\def\mnras{Monthly Not.~R.~Astron.~Soc.}             
\def\pasj{Publ.~Astron.~Soc.~Japan}               
\def\pasp{Publ.~Astron.~Soc.~Pacific}               
\volumenumber{xx}
\publishyear{20xx}
\frompage{\pageref{firstpage}}
\topage{\pageref{finalpage}}

\shorttitle{F. Kemper: Stellar dust production in the Magellanic Clouds}

\title{Stellar dust production and composition in the Magellanic Clouds}
\author{F.~Kemper$^1$}
\affiliation{$^1$Academia Sinica, Institute of Astronomy and Astrophysics, PO Box
23-141, Taipei 10617, Taiwan, R.~O.~C.}
\received{xxxx xx, 2003}\revised{xxxx xx, 2003}\accepted{xxxx xx, 2003}
\abstract{The dust reservoir in the interstellar medium of a galaxy is
  constantly being replenished by dust formed in the stellar winds of
  evolved stars. Due to their vicinity, nearby irregular dwarf
  galaxies the Magellanic Clouds provide an opportunity to obtain a
  global picture of the dust production in galaxies. The Small
  and Large Magellanic Clouds have been mapped with the \emph{Spitzer}
  Space Telescope from 3.6 to 160 $\mu$m, and these wavelengths are especially
  suitable to study thermal dust emission. In addition, a large number
  of individual evolved stars have been targeted for 5-40 $\mu$m
  spectroscopy, revealing the mineralogy of these sources. Here I
  present an overview on the work done on determining the total dust
  production rate in the Large and Small Magellanic Clouds, as well as
  a first attempt at revealing the global composition of the freshly
  produced stardust.  }  
\begin{document}
\label{firstpage}
\maketitle
\copyrighttext{}

\section{Introduction}

Dust is a common ingredient of the interstellar medium (ISM) of
galaxies. In the imaging obtained by optical observatories, such as
the \emph{Hubble Space Telescope}, dust manifests itself as dark lanes
or patches that are obscuring the light from the stars. As such, dust
was often seen as a nuisance, but with the advent of infrared
astronomy, it has become obvious that the absorbed starlight is in
fact re-emitted in the form of thermal emission from dust grains,
thus altering the shape of galaxy spectral energy distributions (SEDs).  The
dust-to-gas ratio in the ISM of galaxies is usually $\lesssim$1\% by
mass for galaxies with Solar metallicities
\citep[e.g.][]{2007ApJ...663..866D}, while typically about 30\% of the
energy emitted by such galaxies is in the form of thermal dust
emission \citep[e.g.][]{2002ApJ...576..159D}.

It is believed that dust in the interstellar medium is formed by
generations of mass-losing evolved stars that gradually increase in
metallicity with chemical galactic evolution
\citep{2003MNRAS.343..427M}.  Dust formation can occur in stars of all
main-sequence masses, but the total dust budget is dominated by the
more numerous Asymptotic Giant Branch (AGB) stars (with
$M_{\mathrm{MS}} < 8 M_\odot$), although Red Supergiants (RSGs;
originating from $M_{\mathrm{MS}} > 8 M_\odot$) are believed to be
important contributors too, based on studies of stellar populations in
the Solar neighbourhood \citep[e.g.][]{W_03_dust}. AGB stars lose mass
at a rate of $10^{-7} - 10^{-6}$ \mdot, and only for a short while,
during the superwind phase, experience mass-loss rates of $10^{-4}$
\mdot\, while RSGs can sustain mass loss at this high rate for longer
\citep[e.g.][]{HO_03_AGBstars,2005A&A...438..273V}.

\section{The Magellanic Clouds}

Due to our vantage point within the Milky Way, and the extinction in
the Galactic plane it is not possible to obtain a full global picture
of the dust production from evolved stars in our own Galaxy. Instead,
a global picture may be obtained from nearby external galaxies, and
the Magellanic Clouds are obvious candidates for such a study. The
Large and Small Magellanic Clouds are irregular dwarf galaxies at a
distance of 50 and 61 kpc respectively
\citep{2008AJ....135..112S,2009AJ....138.1661S}, and both are seen
almost face-on, resulting in a low optical depth along the
line-of-sight, and very little confusion and few chance
superpositions. Unfortunately, due to their vicinity to the much more
massive Milky Way Galaxy, both Magellanic Clouds are tidally
disrupted. Their relatively low metallicities of $Z_\mathrm{LMC} \sim
0.5 \, Z_\odot$ \citep{1982ApJ...252..461D} and $Z_\mathrm{SMC} \sim
0.2 \, Z_\odot$ \citep{Bernard_08_ISM} make the Magellanic Clouds
analogues for Milky Way type galaxies at cosmic times of $\sim 2$ and
$\sim 0.7$ Gyr after the formation of our own Milky Way
\citep{2006MNRAS.366..899N}. The nearest spiral galaxy M31 (Andromeda)
is at 750 kpc significantly further away and to date, the Magellanic
Clouds still provide the best compromise between distance and size to
study the individual galaxy components, particularly stars.

\section{Dust factories}

Thermal emission from dust, particularly warm ($\lesssim 1000$ K)
newly formed circumstellar dust, is most prominent in the near- and
mid-infrared. Thus, surveying a galaxy in the infrared, and
identifying the point sources, will give us a handle on the dust
production by evolved stars. Such surveys have been done in the past
(e.g. DENIS \citet{CVL_00_redgiantbranch}; MSX \citet{EVD_01_LMC}),
but using IRAC \citep{2004ApJS..154...10F} and MIPS
\citep{RYE_04_MIPS}, the first modern infrared surveys of the LMC
\citep{Meixner_06_SAGE} and SMC \citep{Gordon_11_SAGE-SMC} have been
done, cataloguing over 8 million and 2 million infrared point sources
in the LMC and SMC, respectively \citep{Boyer_11_SMC_ES}.  The SAGE
surveys can be supplemented with near-infrared surveys, such as 2MASS
\citep{2006AJ....131.1163S}, the Vista Survey of Magellanic Clouds
\citep[VMC;][]{CCG_11_VMC}, or the AKARI survey of the LMC
\citep{2008PASJ...60S.435I}.

\subsection{Source classification}

 \citet{Blum_06_evolved} used the
2MASS data combined with the IRAC-[3.6] photometry to classify the
dominant categories of dust producing evolved stars in the LMC. The
first incarnation of the SAGE-LMC point source catalogue
\citep{Meixner_06_SAGE} contained about 4 million sources, of which
$\sim$ 820,000 point sources have a $J$ and $[3.6]$ magnitude
measurement, mostly consisting of Red Giant Branch (RGB) stars ($\sim$
650,000), which are not considered significant contributors of dust to
the ISM. Only about 5\% ($\sim 43,000$) of the sources with a reliable
$J$ and $[3.6]$ magnitude are brighter than the tip of the RGB
($[3.6]<11.85$) and thus potentially dust producing. These were
subdivided using colour criteria, into carbon-rich AGB stars ($\sim
7000$), oxygen-rich AGB stars ($\sim 18,000$), (Red) Supergiants
($\sim 1200$) and \emph{extreme} AGB stars ($\sim 1200$). Other,
non-dust-producing sources identified include foreground (Galactic)
and background (extragalactic) objects.

Recently, \citet{Boyer_11_SMC_ES} extended this method of
classification to the SMC by including the SAGE-SMC survey
\citep{Gordon_11_SAGE-SMC}.  They also revised the source counts for
the LMC, using the most recent version of the SAGE-LMC point source
catalogue, which now contains over 2 million sources with both $J$ and
$[3.6]$ band magnitudes, due to improved extraction and calibration
methods\footnote{Both the SAGE-LMC and SAGE-SMC point source catalogues
  are available through the NASA/IPAC Infrared Science Archive {\tt
    http://irsa.ipac.caltech.edu/applications/Gator/}}.  Nevertheless,
the number of point sources brighter than the tip of the RGB has been
revised upward by only $\sim 7$\% to 45,870, and the actual number of
sources in each of the dust forming categories has changed at most by
a similar percentage \citep[See Table 2 of][]{Boyer_11_SMC_ES}.

The classifications performed by \citet{Blum_06_evolved} and
\citet{Boyer_11_SMC_ES}, but also those done by other teams
\citep[e.g.~VMC;][]{CCG_11_VMC} all work with sharply-defined colour
and magnitude cuts to classify the point sources, while in reality the
boundaries may not be that clear, and overlaps between the classes
 in colour-colour and colour-magnitude space may exist.
\citet{2009AJ....138...63M} have developed a more statistical approach
to source classification using magnitudes and colours, based on a
so-called \emph{Nearest Neighbour} ($k$-NN) method.  In this approach
entries in a point source catalogue are assigned probabilities that
they belong to a certain class of objects, based on their distance to,
and the density (in colour-colour or colour-magnitude space) of a
cloud of characteristic objects in that class. Thus, entries in the
point source catalogue can have assigned probabilities to belonging to
more than one object class, making this classification useful mostly
in a statistical sense. The $k$-NN classification will be applied to
all point sources in the SAGE-LMC and SAGE-SMC catalogues (Marengo et
al.~\emph{in prep.}).

\subsection{Stellar dust production rates}

From these classifications, it is possible to estimate the dust
production per source category in the LMC. \citet{Srinivasan_09_excess} focused
on the carbon-rich, oxygen-rich and extreme AGB star
classifications from \citet{Blum_06_evolved} and using the infrared
excess measured at IRAC 8.0 and MIPS 24 $\mu$m found that the overall
dust injection to the interstellar medium of the LMC is $2.74 \times 10^{-5}$
\mdot.  The dust production is dominated by the category of extreme
AGB stars, which are responsible for a dust production rate of $2.36
\times 10^{-5}$ \mdot, against 0.14 and $0.24 \times 10^{-5}$ \mdot\
for oxygen-rich and carbon-rich AGB stars respectively.  An
alternative estimate is presented by \citet{Matsuura_09_dustbudget}
who used representative sources identified by their IRS
\citep{2004ApJS..154...18H} spectra to carve out a classification
scheme in the $[8.0]$ vs.~$[3.6]-[8.0]$ colour-magnitude diagram,
distinguishing between carbon-rich and oxygen-rich AGB stars. Using
empirical mass-loss relations, they were able to estimate dust
production rates of $4.3 \times 10^{-5}$ \mdot\ by carbon-rich AGB
stars, and a lower limit of $4 \times 10^{-6}$ \mdot\ for the
oxygen-rich AGB stars. RSGs were also included in their analysis and
responsible for at least $2 \times 10^{-6}$ \mdot\ of dust production.

To estimate the dust injection in the SMC, the analysis performed by
\citet{Srinivasan_09_excess} for the LMC can be repeated. The colour
cuts lead to the classification of almost 500,000 point sources in the
SMC, with detections in $J$ and $[3.6]$. Of these, $\sim$ 19,000
objects are brighter than the tip of the RGB. About 1700 stars are
classified as carbon-rich AGB stars; about 4000 stars as oxygen-rich
AGB stars, of which 227 as \emph{bright} oxygen-rich AGB stars; a
further 349 stars are \emph{extreme} AGB stars; and around 3300
objects are thought to be RSGs. Thus, using the [8.0] and [24] $\mu$m
excess, the overall dust production rate is thought to be an order of
magnitude lower for the SMC, compared to the LMC \citep{Boyer_12}.

An important potential source of dust is ignored for both the LMC and
the SMC: the dust
production by supernova. This contribution is harder to estimate due to the
incidental occurrence of supernovae, and the difficulties in
establishing the dust production per supernova.  Recently,
\citet{2011Sci...333.1258M} provided a measurement of the dust mass
associated with SN 1987A in the LMC, of 0.4--0.7 \msun, implying that
supernovae can indeed be important dust producers, and providing
an important step forward in establishing the dust
contribution due to supernovae in the ISM of galaxies in the local
universe.

\section{Mineralogy of stardust}

The mineralogical make-up of the freshly formed stellar dust can be
derived from infrared spectroscopy, as many common dust
species have spectral resonances in the mid-infrared, that
can relatively easily be distinguished from each other.

Prior to \emph{Spitzer}, a rare mineralogical analysis of dust
produced by an LMC star was presented by \citet{VWM_99_exgalxsil}, who
detected the presence of crystalline silicates in the luminous blue
variable R71, observed with the Infrared Space Observatory
\citep[ISO;][]{KSA_96_ISO}, using the Short Wavelength Spectrometer
\citep[SWS;][]{GHB_96_SWS}. \citet{TVZ_99_04496} reported the presence
of a silicate feature around a carbon-star in an observation with
limited spectral range, but with a more extensive spectral range it
was later shown that the 'emission feature' was really a lack of molecular
absorption \citep{Speck_06_CStar}.  Further ISO studies
focused on the 10 $\mu$m emission features in M-stars
\citep{DSR_05_lmc}, showing the presence of alumina and amorphous
silicates, with the fraction of silicates increasing as the star
evolves to higher mass loss rates.

With \emph{Spitzer} it became possible to obtain 5--40 $\mu$m
spectroscopy of many point sources in the LMC and SMC. Indeed, many
observing programs were proposed, some targeting just a single or a
few point sources, while other PIs observed a few hundred point
sources with the IRS instrument. At several points during and after
the mission, overviews of the spectral inventory of point sources in
the Magellanic Clouds were given
\citep{2008AJ....136.1221K,2008ApJ...686.1056S,2009AJ....138.1597B,Kemper_10_SAGE-Spec}.
In total $\sim$1000 point sources were observed in the LMC, and
another $\sim$300 in the SMC, covering a wide range of colours and
magnitudes, and thus physical environments.  The spectroscopic
information can help to test the photometric classification schemes
discussed in Sect.~3.1. The 297 staring mode observations performed in
the context of the SAGE-Spec program \citep{Kemper_10_SAGE-Spec}, were
classified with the help of a decision tree
\citep{Woods_11_classification}. The result of this classification was
complementary to and in agreement with the analysis done by
\citet{2009AJ....138.1597B} and \citet{2008AJ....136.1221K}. A full
spectral classification of the IRS staring mode observations in the
LMC (Woods et al.~\emph{in prep.})  and SMC (Ruffle et al.~\emph{in
  prep.}) is in preparation, the results of which will be used to test
the $k$-NN photometric classification (Marengo et al.~\emph{in
  prep.}).

\subsection{Oxygen-rich stars}

Silicates dominate the spectra of both RSGs and oxygen-rich AGB stars
in the LMC.
\citet{DSR_05_lmc} studied a sample of M stars (of AGB and RSG origin)
showing the 10 $\mu$m silicate feature in emission or self-absorption
in the ISO spectroscopy, and found that, while the spectra all show
amorphous silicates, some of them also show the spectral signature of
Al$_2$O$_3$ (alumina). The relative strength of the alumina signature
to the spectrum decreases with increasing mass-loss rate, suggestive
of a covering of the alumina grains by silicate mantles. At the
highest mass-loss rates, the spectra are entirely dominated by
silicate features.

When the IRS spectra of two typical AGB stars in the LMC, one of low
and one of high mass-loss rate, are fitted with the same dust
composition \citep{Sargent_10_oxygen-rich}, it is found that the best
fit is produced by oxygen-deficient amorphous silicates
\citep{OHM_92_cosmicsilicates}, although other amorphous silicate
profiles give good results too. 

Some oxygen-rich stars show the sharp
spectral resonances due to crystalline silicates. In addition to the
presence of crystalline forsterite (Mg$_2$SiO$_4$) around LBV R71
\citep{VWM_99_exgalxsil}, \citet{ZMW_06_SpitzerLMC} showed that
oxygen-rich AGB star IRAS 05003$-$6712 also contains crystalline
forsterite and enstatite (MgSiO$_3$) in its circumstellar shell.  Low
mass-loss rate LMC AGB star HV 2310 also shows crystalline silicates,
particularly around 11.2 $\mu$m, and the crystalline fraction of the
silicates is around 7\% by mass \citep{2006ApJ...638..472S}.  A
comprehensive study of all oxygen-rich AGB stars and RSG in the LMC
and SMC for which IRS spectroscopy is available, shows that most
objects do not show significant crystallinity in their silicates,
although in some cases, the strength of the crystalline silicate
features due to enstatite and forsterite at 23, 28 and 33 $\mu$m can
be up to 10\% above the continuum level (Jones et al.~\emph{in
  prep.}). Given that at the wavelengths of the crystalline silicates
resonances the intrinsic opacities of these crystalline silicates is
much stronger than their amorphous counterparts, these measurements
correspond to crystalline fractions of a few percent, at most. It
needs to be emphasised however, that the lack of crystalline silicate
features in the infrared spectrum does not mean that the crystalline
silicates are absent; in particular in optically thin dust shells,
they may simply be colder than the amorphous silicates
\citep{Kemper_01_xsilvsmdot}.
Thus, I conclude for the purposes of this paper, that oxygen-rich AGB
stars and RSGs in the LMC and SMC predominantly produce silicates, of
an unspecified composition, but that a small fraction of these
silicates may actually be due to crystalline enstatite and forsterite
at the $\sim$ 5\% level by mass.

\subsection{Carbon-rich stars}

\begin{table*}[t] 
\renewcommand{\arraystretch}{1.2}
\vspace{-.3cm}
\caption{Dust production in the LMC by various types of evolved stars, along 
  with the adopted mineralogy in each source category. The bottom row summarises
  the total dust production rate and the mineralogy of the stellar dust
  production for the LMC.  
  Dust production rates are taken from \citet{Matsuura_09_dustbudget} and \citet{Srinivasan_09_excess}. The typical compositions are discussed in Sect.~4 of the main text.}
\vspace{-.1cm}
\begin{center}
\begin{tabular}{ccc} \hline
Object type& Mineralogy & Dust production rate\\
& (mass) & ($\mdot$) \\\hline
O-rich AGB & 95\% am.~sil.; & $0.14\, - >0.4 \times 10^{-5}$  \\ 
& 5\% cryst.~sil.& \\\hline
C-rich AGB & 88\% am.~carbon.; & $0.24 \times 10^{-5}$ \\ 
& 12\% SiC; var.~\% MgS& \\\hline
\emph{extreme} AGB & 88\% am.~carbon;  & $2.36\, -\lesssim 4.3 \times 10^{-5}$ \\ 
& 12\% SiC; var.~\% MgS&\\\hline
RSG  & 95\% am.~sil.;  & $0.2 \times 10^{-5}$ \\ 
 & 5\% cryst.~sil. &\\\hline
total & 77\% am.~carbon; 11\% SiC & $(4\pm1) \times 10^{-5}$\\
& 12\% am.~sil.; $<$1\% cryst.~sil.; var.~MgS &\\ \hline
\end{tabular}
\end{center}
\end{table*}

The mineralogy of carbon-rich AGB stars in the Magellanic Clouds has
been the subject of several studies. Silicon carbide (SiC), amorphous
or graphitic carbon (C) and, in some cases, a solid-state feature
at 30 $\mu$m, previously thought to be due to magnesium sulfide \citep[MgS;][]{HWT_02_MgS,Zhang09},
are seen in the IRS spectra of carbon-rich AGB stars in the LMC
\citep{ZMW_06_SpitzerLMC} and the SMC \citep{2006ApJ...645.1118S}. The 30 $\mu$m feature
is found to be present only in the redder, and hence colder, dust
shells, where the dust temperature drops below 600 K
\citep{ZMW_06_SpitzerLMC}. In the low metallicity environments of the
LMC and the SMC the mass balance between SiC and C in carbon stars
will differ from those found in Galactic carbon stars
\citep[e.g.][]{2006ApJ...645.1118S}, as the required carbon is
produced by the star itself, but the silicon abundance reflects the
composition of the natal cloud. Therefore, less silicon is available
with lower metallicities. Indeed, \citet{LZS_07_carbonstars} confirmed
that the SiC-feature-to-C-continuum ratio in the SMC is lower than in
the LMC, but they also see an evolution with temperature ($[6.4]
-[9.3]$ colour) of the dust shells. While initially the SiC feature
strength increases, with decreasing temperature, the SiC feature
strength eventually drops.  This is also seen by
\citet{Leisenring_08_CStar}, who study the strength of the 30 $\mu$m
emission feature in the same sequence, and extend the sample to bluer
colours, indicative of lower mass-loss rates, particularly for the
Milky Way comparison sample. While \citet{LZS_07_carbonstars} argue
that SiC condenses first, followed by the condensation of C on top of
the SiC, \citet{Leisenring_08_CStar} show that it is not possible with
the LMC or SMC data set to establish the condensation sequence for SiC
and C.  Both studies agree, however, that MgS condenses after SiC and C
\citep{LZS_07_carbonstars,Leisenring_08_CStar}. 
\citet{Srinivasan_10_Cstars} has determined the composition of the dust
shell of a typical, albeit only one, C-rich AGB star, without the MgS
emission feature at 30 $\mu$m, and finds that about 12\% (in a range
of 10--16\%) of the dust by mass is contained in SiC, and that the
remainder is in the form of amorphous carbon.

The class of \emph{extreme} AGB stars is shown by \citet{2008ApJ...688L...9G}
to be made up of carbon-rich objects. In their sample of 13 objects, the
SiC feature is actually seen in absorption, due to the high optical depth.
Only two of the objects studied show the MgS silicate feature,
casting doubt on how common this dust component really is in more evolved
AGB stars. The mass fractions taken up by SiC, C and possibly MgS in \emph{extreme}
AGB stars are not yet determined, but the subject of a future study (Speck
et al.~\emph{in prep.}), and for the purpose of this work I will assume
they are similar to what is derived for lower mass-loss rate C-rich 
AGB stars by \citet{Srinivasan_10_Cstars}.

\section{Results and outlook}

The dust production rates and typical mineralogies for the dust
factories in the LMC are summarised in Table~1. With the analysis so
far, it is possible to derive the mineralogy of the stellar dust
production input into the ISM for the LMC. I find that the majority
($\sim 77$\%) of the freshly produced stardust is in the form of
amorphous or possibly graphitic carbon. Roughly equal amounts of
stardust are taken up by SiC ($\sim 11$\%) and amorphous silicates
($\sim 12$\%). Crystalline silicates, e.g.~enstatite and forsterite,
make up less than 1\% of the dust production by evolved stars. An
unspecified amount of MgS may also be produced by evolved stars.  Apart
from the fact that the 30 $\mu$m feature is only seen in a subset of all C-rich AGB
stars, and therefore hard to include in the analysis of a typical
spectrum, it is also difficult to determine the dust mass contained in
MgS in radiative transfer calculations, because of the lack of optical
and near-infrared optical constants of MgS \citep{HWT_02_MgS}.

Another caveat is the absence of dust production by supernovae in
Table~1. Although a reliable measurement of the dust mass associated
with SN 1987A now exists \citep{2011Sci...333.1258M}, it is difficult
to assess what fraction of that dust is produced by the supernova, and
what fraction is simply processed or heated interstellar dust. In
addition, we need to know how typical this supernova is in terms of
dust production, and what the overall supernova rate is for the LMC
and the SMC, to establish the dust production by supernovae. Finally,
the mineralogy of supernova dust is not well known, with the
mineralogy of supernova ejecta determined by interpreting single
spectral features \citep{ADM_99_SN22um,2009ApJ...700..579R}.

Within the SAGE collaboration, we will continue to improve on the numbers
presented in Table~1, as establishing the integrated composition of
freshly produced stardust is indeed one of the goals of the SAGE-Spec
project \citep{Kemper_10_SAGE-Spec}. The typical compositions derived
by \citet{Srinivasan_10_Cstars} and \citet{Sargent_10_oxygen-rich}
have been used to calculate a grid of SEDs, as a function of dust shell parameters (GRAMS;
\citet{Srinivasan_2011_GRAMS,2011ApJ...728...93S}). The SEDs in this
grid will be fitted to the relevant photometric data in the SAGE-LMC
catalogue, which will then allow us to update in particular the third
column of Table~1, and calculate the total dust production accordingly.

In an alternative approach, the $k$-NN colour classification will allow
for a statistical determination of the numbers in each source category
(Marengo et al.~\emph{in prep.}). Along with a more detailed study of
the mineralogy of a range of objects within source categories
(e.g.~Jones et al.~\emph{in prep.}; Speck et al.~\emph{in prep.}), the
overall mineralogy will be better established as well. This procedure
will also be carried out for the SMC.

\acknowledgments{This research has been supported by the National
  Science Council, under grant code NSC100-2112-M-001-023-MY3.}

\email{F. Kemper (e-mail: ciska@asiaa.sinica.edu.tw)}
\label{finalpage}
\lastpagesettings
\end{document}